\documentclass[fleqn,11pt]{article}
\usepackage{hyperref}
\usepackage{amsfonts,amsmath,amssymb}
\usepackage{latexsym}
\usepackage{mathtools}

\usepackage{amsfonts,amssymb,cite}
\usepackage{graphicx}

\def\noi{\noindent}
\def\jnumber#1#2{\thispagestyle{empty} \noi\unitlength=1mm
    	\begin{picture}(178,10)
            \put(177,15){\llap{\large\it Grav. Cosmol. No.\,#1, #2}}
                    \end{picture}}

\newcommand{\Title}[1]{\noi {{\Large\bf #1}}\\[1ex]}

\def\Aunames#1{\noi{\bf #1}}
\def\au#1{${}^{#1}$}
\def\Addresses#1{\medskip\noi \protect
	\begin{description}\itemsep -3pt {\it #1} \end{description}}
\def\adr#1#2{\item[${}^{#1}$]{\it #2}}

\newcommand{\Abstract}[1]{\vskip 2mm \begin{center}
        \parbox{16.4cm}{\small\noi #1} \end{center}\medskip}

\def\email#1#2{\footnotetext[#1]{e-mail: #2}\addtocounter{footnote}{1}}

\def\nqq{\hspace*{-2em}}

\def\cm{\hspace*{1cm}}

\usepackage{color}

\def\Acknow#1{\subsection*{Acknowledgments} #1}

\def\Jl#1#2{#1 {\bf #2},\ }

\def\ApJ#1 {\Jl{Astroph. J.}{#1}}
\def\CQG#1 {\Jl{Class. Quantum Grav.}{#1}}
\def\DAN#1 {\Jl{Dokl. AN SSSR}{#1}}
\def\GC#1 {\Jl{Grav. Cosmol.}{#1}}
\def\GRG#1 {\Jl{Gen. Rel. Grav.}{#1}}
\def\IJMPD#1 {\Jl{Int. J. Mod. Phys. D}{#1}}
\def\JETF#1 {\Jl{Zh. Eksp. Teor. Fiz.}{#1}}
\def\JETP#1 {\Jl{Sov. Phys. JETP}{#1}}
\def\JHEP#1 {\Jl{JHEP}{#1}}
\def\JMP#1 {\Jl{J. Math. Phys.}{#1}}
\def\NPB#1 {\Jl{Nucl. Phys. B}{#1}}
\def\NP#1 {\Jl{Nucl. Phys.}{#1}}
\def\PLA#1 {\Jl{Phys. Lett. A}{#1}}
\def\PLB#1 {\Jl{Phys. Lett. B}{#1}}
\def\PRD#1 {\Jl{Phys. Rev. D}{#1}}
\def\PRL#1 {\Jl{Phys. Rev. Lett.}{#1}}

\def\lal{&&\nqq {}}

\def\beq{\begin{equation}}
\def\eeq{\end{equation}}
\def\bear{\begin{eqnarray}}
\def\bearr{\begin{eqnarray} \lal}
\def\ear{\end{eqnarray}}
\def\earn{\nonumber \end{eqnarray}}

\def\yy{\\[5pt] {}}

\begin{document}
\twocolumn[
\jnumber{issue}{year}

\Title{Anisotropic Compact Star Model on Finch-Skea Spacetime\yy 
}

\Aunames{Ankita Jangid,\au{a,1}  B. S. Ratanpal,\au{b,2} and K. K. Venkataratnam \au{a,3}}

\Addresses{
\adr a {Department of Physics, Malaviya National Institute of Technology, Jaipur, 302017, India}
\adr b {Department of Applied Mathematics, Faculty of Technology \& Engineering, The Maharaja Sayajirao University of Baroda, Vadodara - 390 001, India}
\adr a {Department of Physics, Malaviya National Institute of Technology, Jaipur, 302017, India}
}


\Abstract
	{In this study, we demonstrate a new anisotropic solution to the Einstein field equations in Finch-Skea spacetime. The physical features of stellar configuration are studied in previous investigations. We create a model that meets all physical plausibility conditions for a variety of stars and plot graphs for \textbf{4U 1820-30}.}
\medskip

] 
\email 1 {ankitajangidphy@gmail.com}
\email 2 {bharatratanpal@gmail.com\\ \cm (Corresponding author)}
\email 3 {ratnamhcu@gmail.com}
\section{Introduction}

In astrophysics, compact stars are usually the endpoint of stellar evolution. These compact objects have incredibly high densities compared to other atomic matter. In astronomy, researchers attracted much attention by studying white dwarfs, neutron stars, and black holes, commonly referred to as compact stars because they have similar structures and characteristics. In compact stars, the matter could be anisotropic. The nature of anisotropy can change the evolution and subsequently, the physical properties of stellar objects. Understanding the non-negligible effects of anisotropy on the stellar object parameters such as mass, pressure, composition, etc., are essential for a viable physical model of a compact star.

In a stellar configuration, anisotropy can happen for a variety of reasons, including the presence of a solid core, type $P$ superfluid, phase transition, mixing of two fluids, the presence of an external field, etc. \cite{1}, \cite{2} and \cite{3} have looked at models of a compact star, uncharged spheres with isotropic pressures. Neutral anisotropic matter was examined by \cite{4}, \cite{5} and \cite{6}, \cite{7}. Among the authors who highlight charged isotropic compact models are \cite{8}, \cite{9}, \cite{10}, \cite{11} and \cite{12}. The general model with charge and anisotropy was analyzed by \cite{13}, \cite{14}  and \cite{15}. \cite{16} did pioneering work on compact objects. They discovered hydrostatic equilibrium after studying static spherically anisotropic symmetry configurations. They conclude that anisotropy may not have been disregarded when calculating the mass and surface redshift of compact stars. \cite{17} argued that matter might be anisotropic when it comes to objects with densities that are significantly higher than nuclear density. According to \cite{18}, superfluid development inside the star may also cause anisotropy in pressure to increase. \cite{19} proposed a number of models with spherically anisotropic distribution at constant densities. The identical task was carried out by  \cite{20}, but with variable densities. \cite{21} explored how anisotropy in pressure affected the mass, structure, and physical characteristics of compact objects and obtained an equation of state relating radial pressure and tangential pressure in various versions of the precise solution. For spherically symmetric static anisotropic stellar configuration, \cite{22}, \cite{23}, \cite{24} all provide a class of accurate solutions to the Einstein field equation. Anisotropy was suggested to be a crucial prerequisite for the dense matter density regime by \cite{25}, \cite{26}. \cite{27} purposefully introduced local anisotropy in the self-gravitating system. In modern research, As proposed by \cite{28} relativistic compact objects, pressure anisotropy cannot be ignored. The investigations by \cite{29} and \cite{30} highlighted a number of intriguing characteristics of exact solutions to the Einstein-Maxwell system for charged anisotropic quark stars. The existence of charge and anisotropy in the star interior has been the subject of numerous recent studies by \cite{31}. Generalized isothermal models were discovered by \cite{32} and superdense models were examined by \cite{33}. The study of \cite{34} contains further new precise solutions for charged anisotropic stars. \cite{35} discovered exact models for charged anisotropic materials with a quadratic equation of state. \cite{36} and \cite{37}, using charged stellar models with Van der Waals and modified Van der Waals equations of state, respectively. 

'In this study, we will produce solutions to Einstein's field equations for anisotropic fluid distributions that are static and spherically symmetric. Because we assume that the matter inside the fluid sphere is uncharged, anisotropy plays a significant role in our concept. We use the compact object 4U 1820-30 to compare our model to observational data.

The work is organized as follows: section \ref{sec:2} contains Einstein field equations and their solutions. The constants of integration are obtained in section \ref{sec:3}. Section \ref{sec:4} contains physical analysis and we conclude the work in section \ref{sec:5}.

\section{ EINSTEIN'S FIELD EQUATIONS AND ITS SOLUTIONS}\label{sec:2}
Consider a static spherically symmetric spacetime metric as
\begin{equation}\label{eq1}
ds^2=  e^{v(r)} dt^2-e^{\lambda(r)} dr^2 -r^2 (d\theta^2 + \sin^2 \theta d\phi^2),
\end{equation}
where $\lambda$and $\nu$ are unknown functions dependent on radial coordinate r. 
 The stress-energy-momentum tensor  for anisotropy matter distribution is of the form
\begin{equation}\label{eq2}
T_{ij}=  (\rho + p)u_i u_j - pg_{ij} + \pi_{ij},
\end{equation}
where $\rho$ and $p$ are energy density and isotopic pressure, respectively, and $u^i$ is the radial 4-velocity vector, $\pi_{ij}$ is anisotropic stress tensor given by

\begin{equation}\label{eq3}
 \pi_{ij} = \sqrt{3}S \  \left [ C_i C_j - \frac{1}{3}(u_i u_j - g_{ij}) \right ],  
\end{equation}
the non-vanishing components of energy-momentum tensor are
\begin{equation}\label{eq4}
 \splitdfrac{T^0_0 = \rho, \ \ \ T^1_1 = - \left ( p + \frac{2S}{\sqrt{3}} \right ),}{T^2_2 = T^3_3 =  \left ( p - \frac{S}{\sqrt{3}} \right )}.
\end{equation}
The relation between radial pressure, tangential pressure, and anisotropic is given by
\begin{equation}\label{eq5}
    p_r= p + \frac{2S}{\sqrt{3}},
\end{equation}
\begin{equation}\label{eq6}
    p_t= p - \frac{S}{\sqrt{3}},
\end{equation}
\begin{equation}\label{eq7}
    p_r - p_t = \sqrt{3}S.
\end{equation}
The Einstien's field equations corresponding to the spacetime metric \eqref{eq1} and the energy-momentum tensor \eqref{eq2} are obtained as (G=c=1)
\begin{equation}\label{eq8}
    8 \pi \rho= \frac{1}{r^2} - e^{-\lambda} \left( \frac{1}{r^2} -\frac{\lambda^{'}}{r}\right),
\end{equation}
\begin{equation}\label{eq9}
    8\pi p_r = e^{-\lambda} \bigg(\frac{1}{r^2}+\frac{v^{'}}{r}\bigg)-\frac{1}{r^2} ,
\end{equation}
\begin{equation}\label{eq10}
    8\pi p_t = \frac{e^{-\lambda}}{4} \bigg[2v^{''}+(v{'}-\lambda{'}) \left(v{'}+\frac{2}{r}\right )\bigg],
\end{equation}
\begin{equation}\label{eq11}
8\pi\sqrt{3}S=( 8\pi p_r-8\pi p_t) .
\end{equation} 

We assume metric potential $ e^{\lambda}$ as,
\begin{equation}\label{eq12}
   e^{\lambda}= 1+\frac{r^2}{R^2} ,
\end{equation}
and
\begin{equation}\label{eq13}
    e^{\nu}= F^2 ,
\end{equation}
then equation \eqref{eq11} is written as
\begin{equation}\label{eq14}
 \splitdfrac{ F{''} - \left(\frac{1+\frac{2r^2}{R^2}}{r(1+\frac{r^2}{R^2})}\right)F{'}-\bigg(\frac{1+\frac{2r^2}{R^2}}{r^2(1+\frac{r^2}{R^2})} +}{(8\pi\sqrt{3}s+\frac{1}{r^2})(1+\frac{r^2}{R^2})\bigg) F = 0 } ,
   \end{equation}

we assume $8\pi\sqrt{3}S$ as
\begin{equation}\label{eq15}
 8\pi\sqrt{3}S=\frac{(\frac{r^2}{R^2}-\frac{4r^4}{R^4})}{4R^2(1+\frac{r^2}{R^2})^3} ,
 \end{equation}
 this choice of $8\pi\sqrt{3}S$ is physically viable as $8\pi\sqrt{3}S =0$ at r = 0.
 
The solution of equation \eqref{eq14} is given by
 \begin{equation}\label{eq16} 
     F= C (1+\frac{r^2}{R^2})^{5/2} + D (1+\frac{r^2}{R^2})^{1/2},
 \end{equation}
 where C and D are constants of integration, this leads
\begin{equation}\label{eq17}
e^{\nu}=\left[C\left(1+\frac{r^2}{R^2}\right)^{5/4}+D\left(1+\frac{r^2}{R^2}\right)^{1/4}\right]^2, 
\end{equation}
   the density, radial pressure and tangential pressure then takes the form,
\begin{equation}\label{eq18}
8\pi \rho=\frac{(r^2 +3 R^2)}{({r^2+ R^2})^2},
\end{equation}
\begin{equation}\label{eq19}
8\pi p_r=\frac{-D r^2 R^2+C(-r^4+3r^2 R^2+4 R^4)}{(r^2+R^2)^{2}(D R^2+C(r^2+R^2))},
\end{equation}
 
\begin{equation}\label{eq20}
    8\pi p_t=\frac{-5 D r^2 R^4+ C(11 r^4 R^2 + 27 r^2 R^4 +16 R^6)}{4(r^3+R^2)^{3}(D R^2 +C(r^2+R^2))}.
\end{equation}
 
 The spacetime metric \eqref{eq1} will take in the following form as
 \begin{equation}\label{eq21}
     d s^2= \splitdfrac{\left[C\left(1+\frac{r^2}{R^2}\right)^{5/4}+D\left(1+\frac{r^2}{R^2}\right)^{1/4}\right]^2dt^2}{ - \left(1+\frac{r^2}{R^2}\right)d r^2-r^2 (d\theta^2 + \sin^2 \theta d\phi^2)}. \end{equation}
The constants of integration, we calculate in the next section.

 \section{DETERMINE CONSTANTS} \label{sec:3}

The Spacetime metric  \eqref{eq21} should continuously match with Schwarzschild exterior spacetime metric
\begin{equation}\label{eq22}
ds^2= \splitdfrac{\bigg(1-\frac{2M}{r}\bigg)dt^2-\left(1-\frac{2M}{r}\right)^{-1}
    dr^2-}{r^2(d\theta^2+\sin^2\theta d\phi^2),}
     \end{equation}
at the boundary of star r = a, and $(p_r)_{(r=a)}$ = 0. These two conditions gives
\begin{figure}
    \centering
    \includegraphics[scale=0.9]{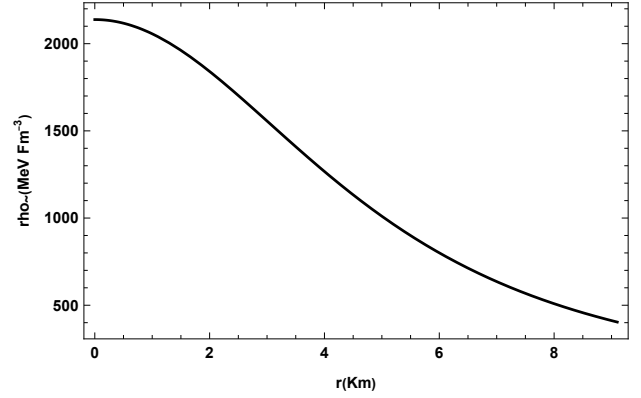}
    \caption{Variation of density $(\rho)$ against the radial parameter $(1MeV fm^{-3}$ = $1.78\times10^{12} g cm^{-3})$}
    \label{fig1}
\end{figure}
\begin{figure}
    \centering
    \includegraphics[scale=0.9]{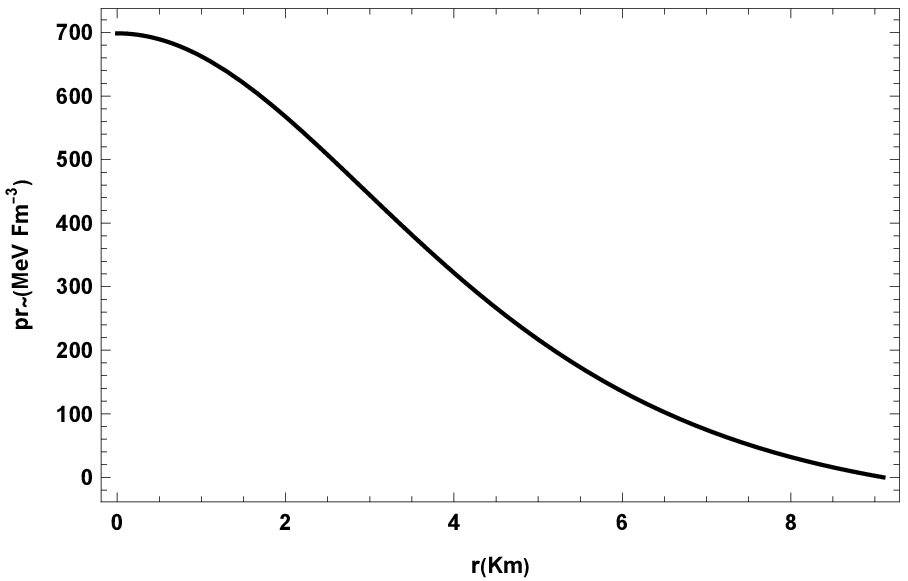}
    \caption{Variation of radial pressure $(p_r)$ against the radial parameter $r$}
    \label{fig2}
\end{figure}
\begin{figure}
    \centering
    \includegraphics[scale=0.9]{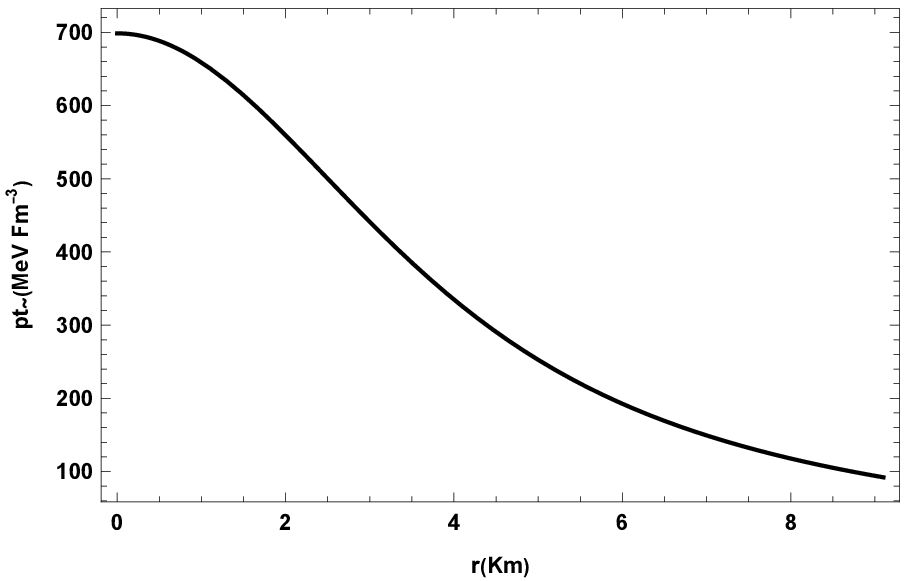}
    \caption{Variation of tangential pressure $(p_t)$ against the radial parameter $r$}
    \label{fig3}
\end{figure}
\begin{figure}
    \centering
    \includegraphics[scale=0.9]{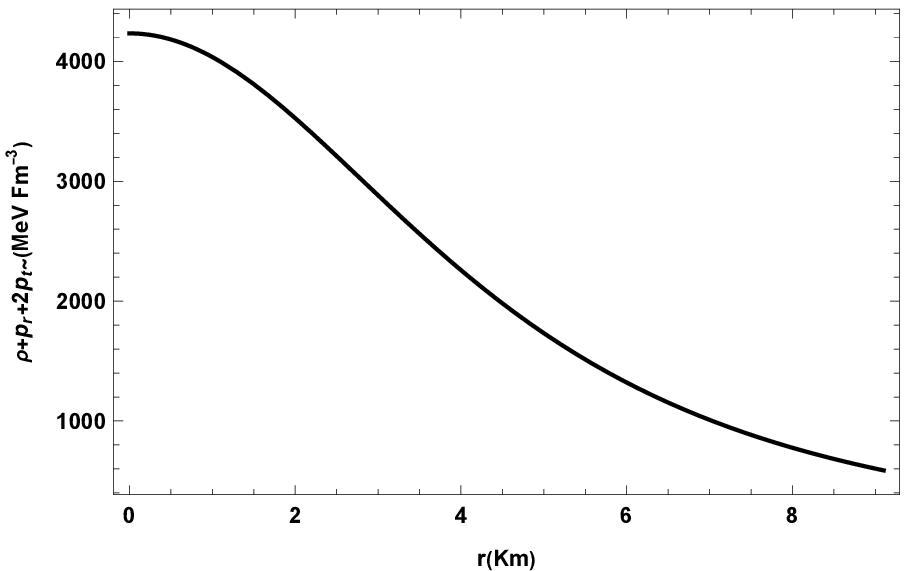}
    \caption{Variation of $( \rho +p_r + 2p_t)$ against the radial parameter $r$}
    \label{fig4}
\end{figure}
\begin{figure}
    \centering
    \includegraphics[scale=0.9]{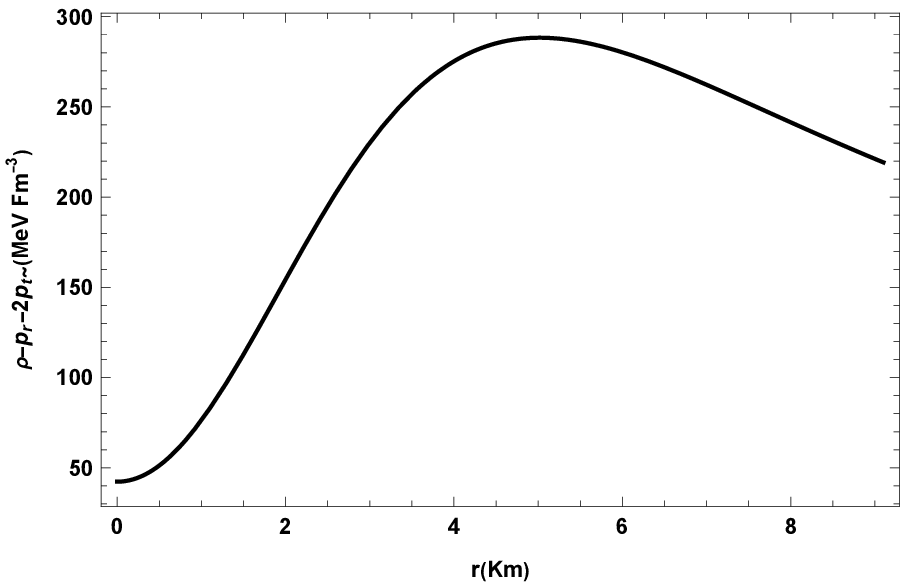}
    \caption{Variation of $(\rho -p_r-2 p_t)$ against the radial parameter $r$}
    \label{fig5}
\end{figure}
\begin{figure}
    \centering
    \includegraphics[scale=0.9]{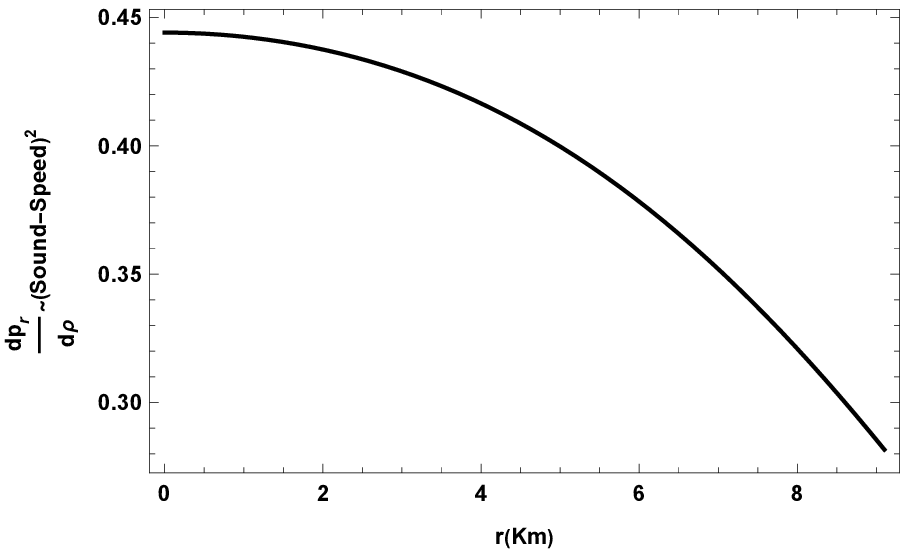}
    \caption{Variation of $\frac{dp_r}{d\rho}$ against the radial parameter $r$}
    \label{fig6}
\end{figure}
\begin{figure}
    \centering
    \includegraphics[scale=0.9]{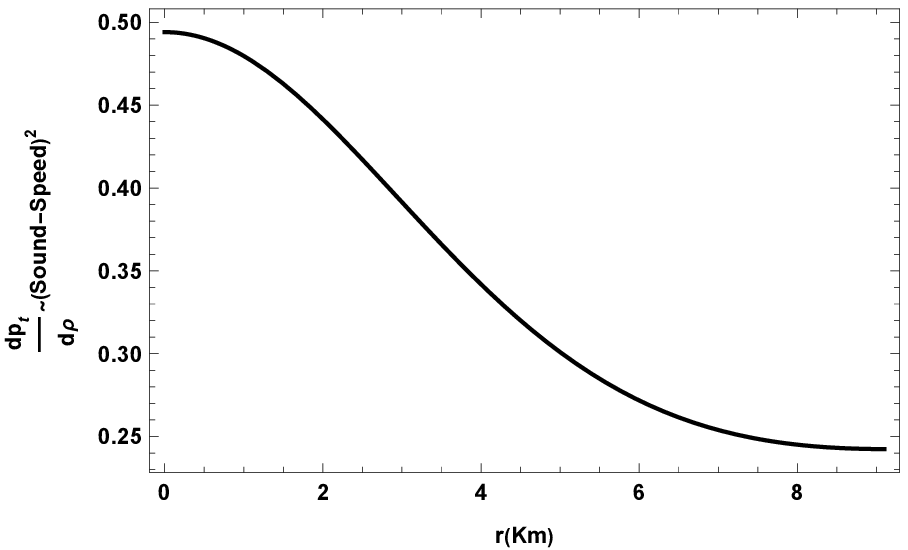}
    \caption{Variation of $\frac{dp_t}{d\rho}$ against the radial parameter $r$}
    \label{fig7}
\end{figure}
\begin{figure}
    \centering
    \includegraphics[scale=0.9]{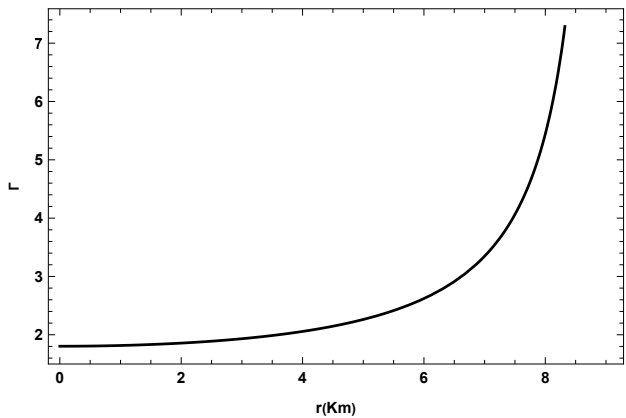}
    \caption{Variation of adiabatic index $(\Gamma)$ against the radial parameter $r$}
    \label{fig8}
\end{figure}
\begin{figure}
    \centering
    \includegraphics[scale=0.9]{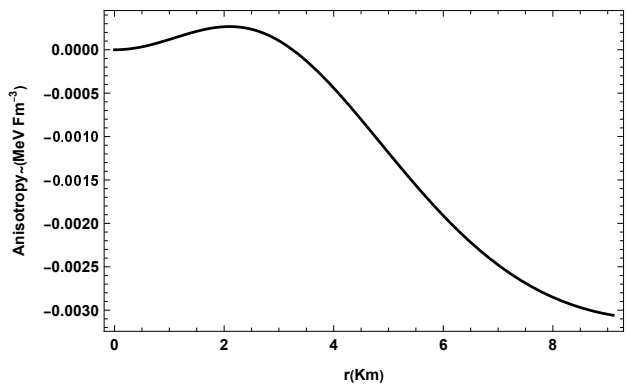}
    \caption{Variation of anisotropic parameter S(r) against the radial parameter $r$}
    \label{fig9}
\end{figure}

  \begin{table*}[!t]
    \centering
    \small\addtolength{\tabcolsep}{-3pt}
    \caption{The estimated and observed values of compact stars}
    \begin{tabular}{|c|c|c|c|c|c|}
    \hline
 \textbf{S.No.} & \textbf{ Star}  & \textbf{$ M_\odot$} & \textbf{R (Chossen)} & \textbf{$\rho_o (Mev/Fm^3)$} & \textbf{$\rho_R(Mev/Fm^3)$ } \\ \hline
  01 & 4U 1820-30 & 1.58 & 6.5 & 2138.24    & 403.491    \\ 
\hline
02  & Her-X 1   & 0.85 & 5.785 & 2698.79   & 509.268 \\
\hline
03 & SMC X-1  & 1.29 & 6.307 & 2270.49  & 428.447 \\
\hline
04  & 4U 1538-52 & 0.87& 5.618 & 2861.75  & 540.019\\
\hline
05  & PSR J1903+327  & 1.667 & 6.735 & 1991.21    & 375.745  \\
\hline
06  &  Vela X-1& 1.77 & 6.828 & 1937.42   & 365.595\\
\hline
 07  & LMC X-4 & 1.04 & 5.929& 2569.68   & 484.904 \\
\hline
08  &PSR J1614-2230 & 1.97 & 6.921 & 1885.78   & 355.852\\ \hline
 \end{tabular}
    \label{tab3}
    \end{table*}

        \begin{equation}\label{eq23}
         M=\frac{a^3}{2 (R^2+ a^2)}, \end{equation}
            \begin{equation}\label{eq24}
          C=\frac{ a^2}{(1+\frac{a^2}{R^2})^{3/4}[a^2(1+\frac{a^2}{R^2})+R^2(\frac{-a^4}{R^4}+\frac{3 a^2}{R^2}+4)]} ,
     \end{equation}
      \begin{equation}\label{eq25}
          D=\frac{4R^2-a^2}{4R^2(1+\frac{a^2}{R^2})^{3/4} }. 
      \end{equation}

     Substituting the value of $ C$ and $D$ in eq. \eqref{eq18}, \eqref{eq19} and \eqref{eq20}, we get
      \begin{equation}\label{eq26}
        8\pi   \rho=\frac{(r^2 +3 R^2)}{({r^2+ R^2})^2},
     \end{equation}
     \begin{equation}\label{eq27}
        8\pi   p_r=\frac{a^4 r^2-4 r^2 R^4 -a^2(r^4-4 R^4)}{(r^2+R^2)^{2}(-a^4+4 R^4 +a^2(r^2+4 R^2))},
      \end{equation}
      
      \begin{equation}\label{eq28}
          8\pi   p_t=\frac{\splitdfrac{R^2(5a^4r^2-20r^2R^4+a^2}{(11r^4+12r^2R^2+16R^4))}}{(r^2+R^2)^3(-a^4+4 R^4+a^2(r^2+4 R^2))} .
      \end{equation}
     \section{Physical analysis} \label{sec:4}
     The physically plausible model should satisfy the following conditions
\begin{itemize}
    \item[(i)] $\rho (r), p_r(r), p_t(r) \geq 0, for \ \ 0 \leq r \leq R$
    \item[(ii)] $\frac{d \rho}{dr}, \frac{d p_r}{d r}, \frac{d p_t}{d r} < 0, for \ \ 0 \leq r \leq R$ 
      \item[(iii)]$\rho + p_r  > 0, \ \  for \ \ 0 \leq r \leq R$
    
     \item[(iv)]$\rho + p_r + 2 p_t  \geq 0, \ \  for \ \ 0 \leq r \leq R$

    \item[(v)]$\rho - p_r -2 p_t \geq for \ \ 0 \leq r \leq R$
   
    \item[(vi)] $0 < \frac{d p_r}{d\rho} < 1;\; 0 < \frac{d p_t}{d\rho} < 1,\; for \ \ 0 \leq r \leq R$
\end{itemize}      
For a physically acceptable model, all the physical plausibility conditions are satisfied for stars in table\ref{tab3}. We have demonstrated graphs for a star $ 4U 1820-30 $. The radial pressure, the tangential pressure, and the density should be positive throughout the mass distribution. Fig(s). \ref{fig1}, \ref{fig2} and \ref{fig3} shows that these conditions are satisfied throughout the distribution and decrease with mass distribution. $\rho \geq 0$ is called the null energy condition, which is also satisfied. The third condition is the weak energy condition, which is $\rho + p_r > 0$ fulfills the requirement. Further, the strong energy condition, $ \rho + p_r + 2 p_t \geq 0 $ and the trace energy condition $\rho - p_r - 2 p_t \geq 0 $ are also satisfied throughout
the distribution, as shown in Fig(s). \ref{fig4} and \ref{fig5}. The fifth condition is called the causality condition, followed by star 4U 1820-30 shown in Fig(s)  \ref{fig6} and \ref{fig7}.

It is clear from figure \ref{fig8} that adiabatic index $\Gamma> \frac{4}{3}$.
The variation of anisotropy is shown in Figure \ref{fig9}, and the computed value of central density and surface density for a particular choice of R are shown in table \ref{tab3}.

\section{Conclusion}\label{sec:5}
In this work, we have obtained the exact non-singular solution of Einstein's field equations.
The salient features of the model are that all the physical plausibility conditions are satisfied for stars are shown in table \ref{tab3}, and we have plotted graphs for star 4U 1820-30.   According to \cite{38}, adiabatic index $\Gamma>\frac{4}{3}$, the model is potentially  stable. The data demonstrate that the model satisfies all criteria for physical requirements for 4U 1820-30.

\Acknow{AJ and BSR are grateful to IUCAA, Pune, for their hospitality and the workspace, they were given while working on this project.
}



\small
\begin{thebibliography}{99}
 \bibitem{1}  
 M H. Murad and N. Pant, ``A class of exact isotropic solutions of Einstein’s equations and relativistic stellar models in general relativity,'' Astrophysics and Space Science {\bf 350}, 349-359 (2014). 
 \bibitem{2}  
 M. K. Mak and T.Harko, ``Relativistic compact objects in isotropic coordinates,'' Pramana {\bf 65}, 185-192 (2005). 
  
 \bibitem{3}  
R. Sharma, S. Karmakar and S. Mukherjee, ``Maximum mass of a class of cold compact stars,'' International Journal of Modern Physics D  {\bf 15}, 405-418 (2006).

\bibitem{4}  
B. C. Paul, P. K. Chattopadhyay and S. Karmakar and R. Tikekar,  ``Relativistic strange stars with anisotropy,'' Modern Physics Letters A {\bf 26}, 575-587 (2011).
\bibitem{5}  
T. Harko and M. K. Mak, ``Anisotropic relativistic stellar models,'' Annalen der Physik {\bf 11}, 3-13 (2002).
\bibitem{6}  
F. Rahaman and others, ``Strange stars in Krori--Barua space-time,'' The European Physical Journal C {\bf 72}, 1-9 (2012).

\bibitem{7}  
M. Kalam, A. A. Usmani and others, ``A relativistic model for strange quark star,'' International Journal of Theoretical Physics {\bf 52}, 3319-3328 (2013).
\bibitem{8}  
S. K. Maurya, Y. K. Gupta, ``A family of well-behaved charge analogues of a well-behaved neutral solution in general relativity,'' Astrophysics and Space Science {\bf 332}, 481-490 (2011).
\bibitem{9}  
S. K. Maurya, Y. K. Gupta, ``A class of charged analogues of Durgapal and Fuloria superdense star,'' Astrophysics and Space Science {\bf 331}, 135-144 (2011).
\bibitem{10}  
R. P. Negreiros and F. Weber, ``Electrically charged strange quark stars,'' Physical Review D {\bf 80}, 083006 (2009).
 \bibitem{11}  
 M. H. Murad and  S. Fatema, ``A family of well-behaved charge analogues of Durgapal’s perfect fluid exact solution in general relativity II,'' Astrophysics and space science {\bf 344}, 69-78 (2013). 
 \bibitem{12}  
N. Bijalwan, ``Charged analogues of Schwarzschild interior solution in terms of pressure,'' Astrophysics and Space Science {\bf 336}, 413-418 (2011).
\bibitem{13}  
M. Esculpi and E. Aloma, ``Conformal anisotropic relativistic charged fluid spheres with a linear equation of state,'' The European Physical Journal C {\bf 67}, 521-532 (2010).
\bibitem{14}  
 P. Mafa Takisa and  S. D. Maharaj, ``Compact models with regular charge distributions,'' Astrophysics and Space Science {\bf 343}, 569-577 (2013).
 \bibitem{15}  
F. Rahaman and others, ``Anisotropic strange star with de Sitter spacetime,'' The European Physical Journal C {\bf 72}, 1-7 (2012).
\bibitem{16}  
R. L. Bowers and E. P. T. Liang, ``Anisotropic spheres in general relativity,'' The Astrophysical Journal {\bf 188}, 657 (1974).
\bibitem{17}  
M. Ruderman, ``Pulsars: structure and dynamics,'' Annual Review of Astronomy and Astrophysics {\bf 10}, 427-476 (1972).
\bibitem{18}  
R. Kippenhahn and A. Weigert, ``Stellar structure and evolution,'' Astronomy and Astrophysics Library   {\bf 192}, 1027 (1990).
\bibitem{19}  
S. D. Maharaj and R. Maartens, ``Anisotropic spheres with uniform energy density in general relativity,'' General relativity and gravitation  {\bf 21}, 899-905 (1989).
\bibitem{20}  
M. K. Gokhroo and A. L. Mehra, ``Anisotropic spheres with variable energy density in general relativity,'' General relativity and gravitation  {\bf 26}, 75-84 (1994).
\bibitem{21}  
K. Dev and M. Gleiser, ``Anisotropic stars: exact solutions,'' General relativity and gravitation  {\bf 34}, 1793-1818 (2002).
\bibitem{22}  
K. N. Singh, N. Pradhan `` A New Charged Anisotropic Compact Star Model in General Relativity,'' int. J. Theor. Phys.  {\bf 54}, 3408 (2015).
\bibitem{23}  
K. N. Singh, N. Pant ``Charged anisotropic superdense stars with constant stability factor,'' Astrophysics and Space Science  {\bf 358}, 1-13 (2015).
\bibitem{24}  
K. N. Singh, N. Pant ``Singularity free charged anisotropic solutions of Einstein-Maxwell field equations in general relativity,'' Indian Journal of Physics {\bf 90}, 843-851 (2016).
\bibitem{25}  
M. K. Mak and T. Harko, ``Anisotropic stars in general relativity,'' Proceedings of the Royal Society of London. Series A: Mathematical, Physical and Engineering Sciences {\bf 459}, 393-408 (2003).
\bibitem{26}  
R. Sharma, S. Mukherjee and S.D. Maharaj, ``General solution for a class of static charged spheres, '' General Relativity and Gravitation {\bf 33}, 999-1009 (2001).
\bibitem{27}  
L. Herrera and N. O. Santos,  ``Local anisotropy in self-gravitating systems,'' Physics Reports {\bf 286}, 53-130 (1997).
\bibitem{28}  
L. Herrera, ``Stability of the isotropic pressure condition,'' Physical Review D {\bf 101}, 104024 (2020).
\bibitem{29}  
S. D. Maharaj, J. M. Sunzu and S. Ray,  ``Some simple models for quark stars,'' The European Physical Journal Plus {\bf 129}, 1-10 (2014).
\bibitem{30}  
J. M. Sunzu, S. D.  Maharaj and S. Ray,  ``Charged anisotropic models for quark stars,'' Astrophysics and Space Science {\bf 352}, 719-727 (2014).
\bibitem{31}  
S. D. Maharaj and P. Mafa Takisa,  ``Regular models with quadratic equation of state,'' General Relativity and Gravitation {\bf 44}, 1419-1432 (2012).
\bibitem{32}  
S. D. Maharaj and S. Thirukkanesh,  ``Generalized isothermal models with strange equation of state,'' Pramana {\bf 72}, 481-494 (2009).
\bibitem{33}  
S. K. Maurya and Y. K. Gupta,  ``A family of anisotropic super-dense star models using a space-time describing charged perfect fluid distributions,'' Physica Scripta {\bf 86}, 025009 (2012).
 \bibitem{34}  
    P. Takisa and S. D. Maharaj, ``Some charged polytropic models,'' General Relativity and Gravitation {\bf 45}, 1951-1969 (2013).
\bibitem{35}  
   T. Feroze and A. A. Siddiqui, ``Charged anisotropic matter with a quadratic equation of state,'' General Relativity and Gravitation {\bf 43}, 1025-1035 (2011).
  
  \bibitem{36}  
  M. Malaver, ``Regular model for a quark star with Van der Waals modified equation of state,'' World Applied Programming {\bf 3}, 309-313 (2013).

 \bibitem{37}  
  M. Malaver, ``Analytical model for charged polytropic stars with Van der Waals Modified Equation of State,'' American Journal of Astronomy and Astrophysics {\bf 1}, 41-46 (2013).
  
\bibitem{38}  
B. S. Ratanpal,  ``Cracking and stability of non-rotating relativistic spheres with anisotropic internal stresses,'' IOP Science {\bf 1}, 025207 (2020).

\end{thebibliography}
\end{document}